\documentclass[sigconf]{acmart}

\AtBeginDocument{%
  \providecommand\BibTeX{{%
    \normalfont B\kern-0.5em{\scshape i\kern-0.25em b}\kern-0.8em\TeX}}}

\setcopyright{acmcopyright}
\copyrightyear{2018}
\acmYear{2018}
\acmDOI{XXXXXXX.XXXXXXX}

\acmConference[Conference acronym 'XX]{Make sure to enter the correct
  conference title from your rights confirmation emai}{June 03--05,
  2018}{Woodstock, NY}
\acmPrice{15.00}
\acmISBN{978-1-4503-XXXX-X/18/06}

\begin{document}

\title{Efficient Transfer Learning Framework for Cross-Domain Click-Through Rate Prediction}

\author{Qi Liu$^*$}
\affiliation{%
  \institution{University of Science and Technology of China}
  \city{Hefei}
  \country{China}
}
\email{qiliu67@mail.ustc.edu.cn}

\author{Xingyuan Tang$^*$}
\affiliation{%
  \institution{Meituan}
  \city{Beijing}
  \country{China}
}
\email{tangxingyuan@meituan.com}

\author{Jianqiang Huang}
\affiliation{%
  \institution{Meituan}
  \city{Beijing}
  \country{China}
}
\email{huangjianqiang@meituan.com}

\author{Xiangqian Yu}
\affiliation{%
  \institution{Tsinghua University}
  \city{Beijing}
  \country{China}
}
\email{yxq21@mails.tsinghua.edu.cn}

\author{Haoran Jin}
\affiliation{%
  \institution{University of Science and Technology of China}
  \city{Hefei}
  \country{China}
}
\email{haoranjin@mail.ustc.edu.cn}

\author{Jin Chen}
\affiliation{%
  \institution{University of Electronic Science and Technology of China}
  \city{Chengdu}
  \country{China}
}
\email{chenjin@std.uestc.edu.cn}

\author{Yuanhao Pu}
\affiliation{%
  \institution{University of Science and Technology of China}
  \city{Hefei}
  \country{China}
}
\email{puyuanhao@mail.ustc.edu.cn}

\author{Defu Lian}
\affiliation{%
  \institution{University of Science and Technology of China}
  \city{Hefei}
  \country{China}
}
\email{liandefu@ustc.edu.cn}

\author{Tan Qu}
\affiliation{%
  \institution{Meituan}
  \city{Beijing}
  \country{China}
}
\email{qutan@meituan.com}

\author{Zhe Wang}
\affiliation{%
  \institution{Meituan}
  \city{Beijing}
  \country{China}
}
\email{wangzhe65@meituan.com}

\author{Jia Cheng}
\affiliation{%
  \institution{Meituan}
  \city{Beijing}
  \country{China}
}
\email{jia.cheng.sh@meituan.com}

\author{Jun Lei}
\affiliation{%
  \institution{Meituan}
  \city{Beijing}
  \country{China}
}
\email{leijun@meituan.com}

\renewcommand{\shortauthors}{Trovato and Tobin, et al.}

\begin{abstract}
Natural content and advertisement coexist in industrial recommendation systems but differ in data distribution. Concretely, traffic related to the advertisement is considerably sparser compared to that of natural content, which motivates the development of transferring knowledge from the richer source natural content domain to the sparser advertising domain. Previous efforts have either focused on pre-training with source data and fine-tuning with target advertising data or on regarding hidden representations of pre-training models as knowledge that acts as extra input of the target advertisement model, but these approaches face significant challenges. The challenges include the inefficiencies arising from the management of extensive source data and the problem of 'catastrophic forgetting' that results from the CTR model's daily updating. To this end, we propose a novel tri-level asynchronous framework, i.e., Efficient Transfer Learning Framework for Cross-Domain Click-Through Rate Prediction (E-CDCTR), to transfer comprehensive knowledge of natural content to advertisement CTR models. This framework consists of three key components: Tiny Pre-training Model ((TPM), which trains a tiny CTR model with several basic features on long-term natural data; Complete Pre-training Model (CPM), which trains a CTR model holding network structure and input features the same as target advertisement on short-term natural data; Advertisement CTR model (A-CTR), which derives its parameter initialization from CPM together with multiple historical embeddings from TPM as extra feature and then fine-tunes on advertisement data. These three models are updated monthly, weekly, and daily respectively, ensuring efficient training and deployment under extensive data. TPM provides richer representations of user and item for both the CPM and A-CTR, effectively alleviating the forgetting problem inherent in the daily updates. CPM further enhances the advertisement model by providing knowledgeable initialization, thereby alleviating the data sparsity challenges typically encountered by advertising CTR models. Such a tri-level cross-domain transfer learning framework offers an efficient solution to address both data sparsity and `catastrophic forgetting', yielding remarkable improvements. Extensive experiments conducted on the industrial dataset demonstrate the effectiveness and efficiency of E-CDCTR. Moreover, the A/B test results indicate that E-CDCTR enhances CTR and Revenue per Mille (RPM) by 2.9\% and 2.1\%, respectively in Meituan's online advertising system.
\end{abstract}




\received{20 February 2007}
\received[revised]{12 March 2009}
\received[accepted]{5 June 2009}

\maketitle
\def\thefootnote{*}\footnotetext{The two authors contributed equally to this work.}

\section{Introduction}
\begin{figure}[htb]
    \centering
    \includegraphics[width=0.85\linewidth]{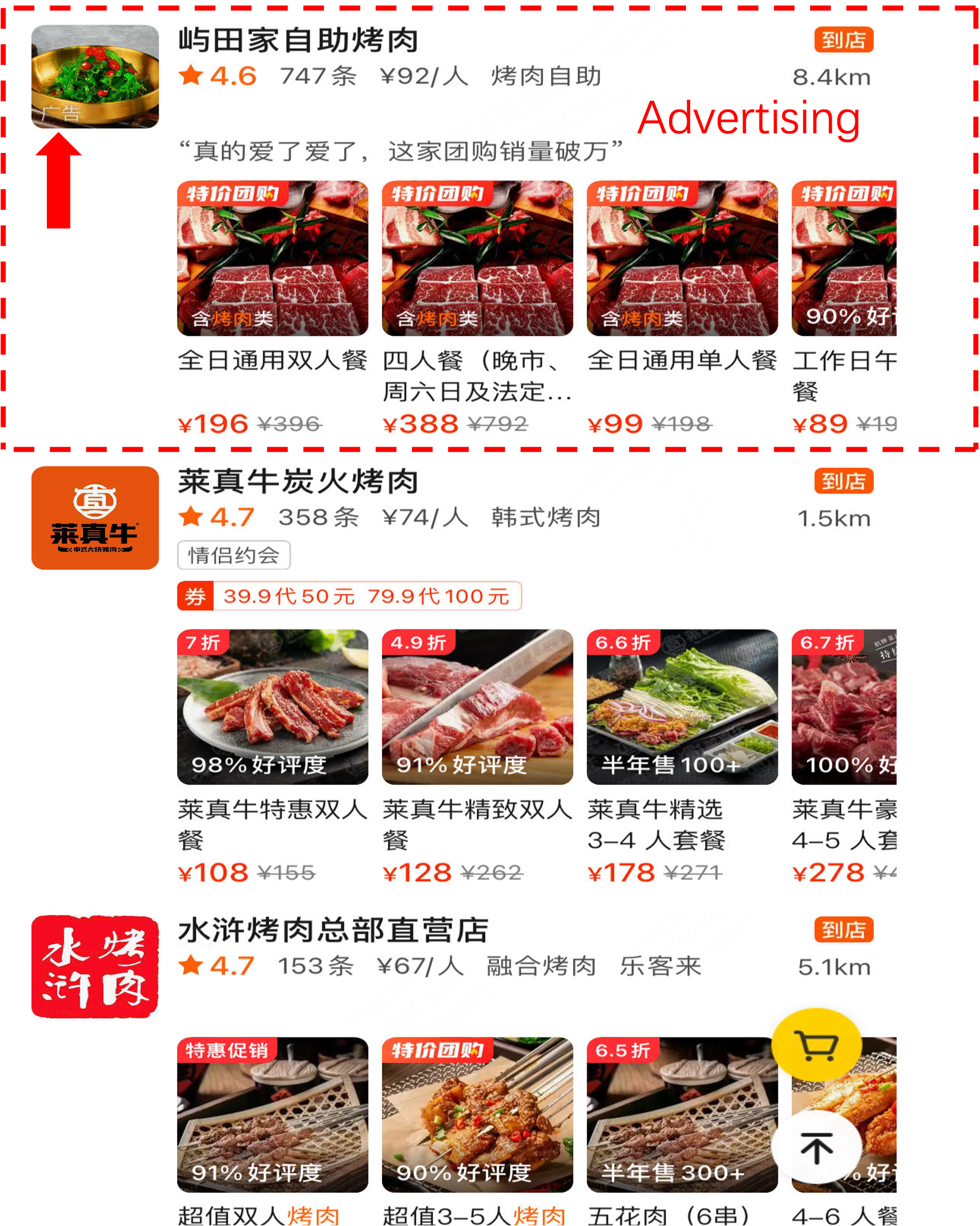}
    \caption{The illustration of blending between natural content and advertisement in the Meituan platform. The merchant bounded by a dashed red rectangle is the advertisement and the other two are the natural content.}
    \label{fig:natural_ad}
\end{figure}

Industrial recommendation systems often feature a blend of natural content and advertisement as shown in Figure~\ref{fig:natural_ad}, where items are presented to diverse users with the aim of enhancing user engagement and improving online revenue~\cite{kong2023lovf}. Among the exposed items, the sellers who want to promote their products may invest in advertising for more impressions~\cite{chen2022hierarchically, wang2019learning}. Consequently, the natural content and advertisements coexist within industrial recommendation systems. While these items and users occupy the same screen, the data distribution differs significantly between them, including the volume of user feedback, the types of user behaviors, and the distribution of items. It is logical to consider these as two different scenarios, where the source domain, enriched by extensive user feedback, serves as a reservoir of valuable insights for the target domain of advertising with the aim of alleviating the problem of data sparsity~\cite{liu2023continual}.

Existing cross-domain recommendation methods~\cite{zhu2021cross} can be widely categorized into two groups. One leverages the multi-task learning method to jointly improve the overall metrics~\cite{li2020ddtcdr,ma2018modeling,ouyang2020minet,sheng2021one}. This method employs shared-bottom representations to establish connections between different domains using mixed data as input. This method often incurs the negative transfer problem across domains~\cite{zhang2021survey}. Furthermore, the source domain, which primarily features natural content, often exhibits a significantly larger dataset, with a scale several times greater than that of the target advertisement data. This discrepancy can exacerbate the imbalance between the two scenarios and result in the suboptimal learning of the target domain with much more sparse user feedback. The second one frequently adopts the pre-training and fine-tuning paradigm to fit in the knowledge transfer framework~\cite{chen2021user,liu2023continual}, where a pre-trained model is trained on the large-scale source domain and subsequently fine-tuned on the target domain to enhance model performance. Notably, in this paradigm, the fine-tuning process for the target domain relies solely on the data from the target domain and is independent of the initial large-scale source domain training. Separating the two phases contributes to an efficient and practical deployment strategy. Moreover, the objective of the fine-tuned model is meticulously tailored to the specific characteristics of the target domain, resulting in superior performance when compared to the joint training approach utilizing mixed data across domains.


However, the pre-training and fine-tuning paradigm for transferring source data from large-scale natural content to the target advertising data still faces the following issues: 
(1) The volume of the data in the source domain, primarily consisting of natural content, is notably substantial, which simultaneously introduces challenges in terms of the efficiency of model training. The extensive natural content, though valuable for enhancing the capabilities of models, demands an unacceptable amount of time for training. When considering a scenario where six months of data is utilized for training, the training process would necessitate two days. The extended time cost poses a significant concern given the need for models to remain adaptable to users' interest shifts in online recommendation systems. (2) The data for the model's daily updating is changing in a sliding window manner in order to strike a balance between training efficiency and capturing shift of data distribution. Concretely, the daily serving model is trained using the last dozens of days' data, which will give rise to the catastrophic forgetting problem~\cite{kirkpatrick2017overcoming}. Regrettably, this focus on recent data causes the model only to capture the near-term personalization characteristic between the user and the item while losing the long-term characteristic. This phenomenon presents a significant challenge as it disrupts the retention of a comprehensive, enduring understanding of user preferences, which is vital for delivering accurate recommendations. 

To this end, we propose \textbf{E}fficient Transfer Learning Framework for \textbf{C}ross-\textbf{D}omain \textbf{C}lick-\textbf{T}hrough \textbf{R}ate (\textbf{E-CDCTR}) prediction to facilitate the transfer of knowledge from natural content to advertising models. This framework is characterized by a \textit{triple-level} asynchronous structure, comprising three components. To mitigate the inefficiency in pre-training caused by the substantial volume of natural content data, we first derive the logs from the extensive long-term natural content data to establish two asynchronous pre-training models: Tiny Pre-training Model (TPM) which trains a tiny CTR model with a few basic features on long-term natural data; and Complete Pre-training Model (CPM), designed to leverage less natural data with complete features on a large CTR model. This departure from the conventional single, comprehensive pre-training model represents a notable shift in our methodology. Specifically, TPM receives less frequent updates compared to CPM and CPM provides parameters for the initialization of the advertisement CTR model (A-CTR). This achieves efficiency in pre-training under the huge volume of natural content, since only CPM with recent data requires frequent updates, thereby reducing the overall training time. Additionally, TPM provides long-term personalization information to CPM and A-CTR in the form of multiple user and item embeddings, addressing concerns of forgetting caused by sliding window-based daily updating.  

The main contribution can be summarized in the following folds:
\begin{itemize}
    \item E-CDCTR achieves efficient knowledge transfer through the design of a tri-level asynchronous pre-training framework, reducing the time cost due to a large amount of source data.
    \item The Tiny Pre-training Model (TPM), which pre-trains a tiny CTR model with a few features on long-term natural data, is proposed to record the historical personalization information for the downstream CTR models, alleviating the forgetting problem caused by daily updating.
    \item We also propose the Complete Pre-training Model (CPM) for natural data with short-term data, whose parameters are adopted to initialize subsequent advertisement model parameters. This mitigates the problem of insufficient model convergence caused by sparse advertisement data.
    \item Experiments conducted on the industrial datasets demonstrate the effectiveness of the proposed E-CDCTR for advertisement, and the online A/B test indicates E-CDCTR achieves a relative improvement of 2.9\% and 2.1\% in terms of CTR and Revenue Per Mile (RPM) respectively.  
\end{itemize}

\section{Related Works}
The most relevant research fields of our work are deep CTR modeling and cross-domain knowledge transfer for CTR. In this section, we give a brief introduction.

\subsection{Deep CTR Prediction}
The prediction of Click-Through Rates (CTR) has consistently held a prominent position in the realm of recommendation Systems research~\cite{zhang2021deep}. Early CTR prediction methods predominantly revolved around low-order feature interactions. However, recent advances in deep learning have ushered in remarkable progress in CTR prediction techniques. Wide\&Deep~\cite{cheng2016wide} was first proposed by Google, which combines a linear model to memorize feature interactions with a deep neural network to enhance generalization. Its remarkable improvements over the online recommendations bring in the furnished utilization of deep models. DeepFM~\cite{guo2017deepfm} further replaces the linear model with Factorization Machines (FM) to emphasize second-order feature interactions. xDeepFM~\cite{lian2018xdeepfm} introduces the Compressed Interaction Network to explicitly capture high-order feature interactions while DCN~\cite{wang2017deep} applies a cross-vector network to autonomously learn informative feature interactions.
Another line goes through the modeling of user historical behaviors. This approach focuses on extracting a user's interests from their historical behaviors to refine preference estimation accuracy. Due to online latency constraints, most existing methods are tailored for truncated short user behavior sequences, encompassing only the user's immediate interests. DIN~\cite{zhou2018deep} stands out by introducing attention mechanisms between candidate items and the behavior sequence. This attention mechanism emphasizes target-relevant behaviors while suppressing target-irrelevant ones to extract interest. DIEN~\cite{zhou2019deep} takes the concept further by incorporating a two-layer Gated Recurrent Unit (GRU)~\cite{chung2014empirical} to model temporal shifts and mine interests at an interest-level granularity. Works~\cite{pi2020search,chang2023twin} devote to extracting long-term interest from the user's life-long behavior sequence by taking an advanced index method to reduce cost.

\subsection{Cross-Domain Knowledge Transfer for CTR}
To alleviate the data sparsity problem, cross-domain knowledge transfer for CTR prediction has been explored in recent years. The first type of work focuses on the co-training paradigm based on the share-bottom architecture to bring useful knowledge from the source domain into the target domain. CoNet~\cite{hu2018conet} combines collaborative filtering with neural networks. It utilizes the neural network to achieve symmetrical knowledge transfer between the source and target domain based on shared user embeddings. MiNet~\cite{ouyang2020minet} tries to perform cross-domain knowledge transfer through the interest extracted from the user behavior sequence. It takes the interest extracted from the source domain to enhance the user's understanding of the target domain. DARec~\cite{yuan2019darec} takes the domain adversarial neural networks to learn the shared rating patterns from the source and target domain by introducing a domain classifier. AutoFT~\cite{yang2021autoft} applies reinforcement learning to decide which parameters should be migrated to the target domain, which aims at alleviating the potential negative knowledge transfer toward the target domain. Recent work KEEP~\cite{zhang2022keep} is the most similar paradigm of knowledge transfer with us. It stores the user/item embedding at the end of training on the source domain dataset and plugs the knowledge into the last layer of Multi-Layer Perceptron (MLP) when training the CTR model of the target domain. CTNet~\cite{liu2023continual} achieves cross-domain knowledge transfer by distilling the hidden representations of the tiny CTR model pre-trained on the source domain to the big CTR model of the target domain. However, none of the existing methods achieve sufficient transfer of knowledge. KEEP only keeps the user/item embedding of the final model of the source domain. Besides, it uses fewer features and a tiny model for training efficiency, which results in the knowledge contained in the user/item embedding is not comprehensive. CTNet shares the same shortcomings. The parameter initialization paradigm of AutoFT can only utilize small amounts of source domain data due to the training cost. 

\section{Method}

\begin{figure*}[htb]
    \centering
    \includegraphics[width=1.0\linewidth]{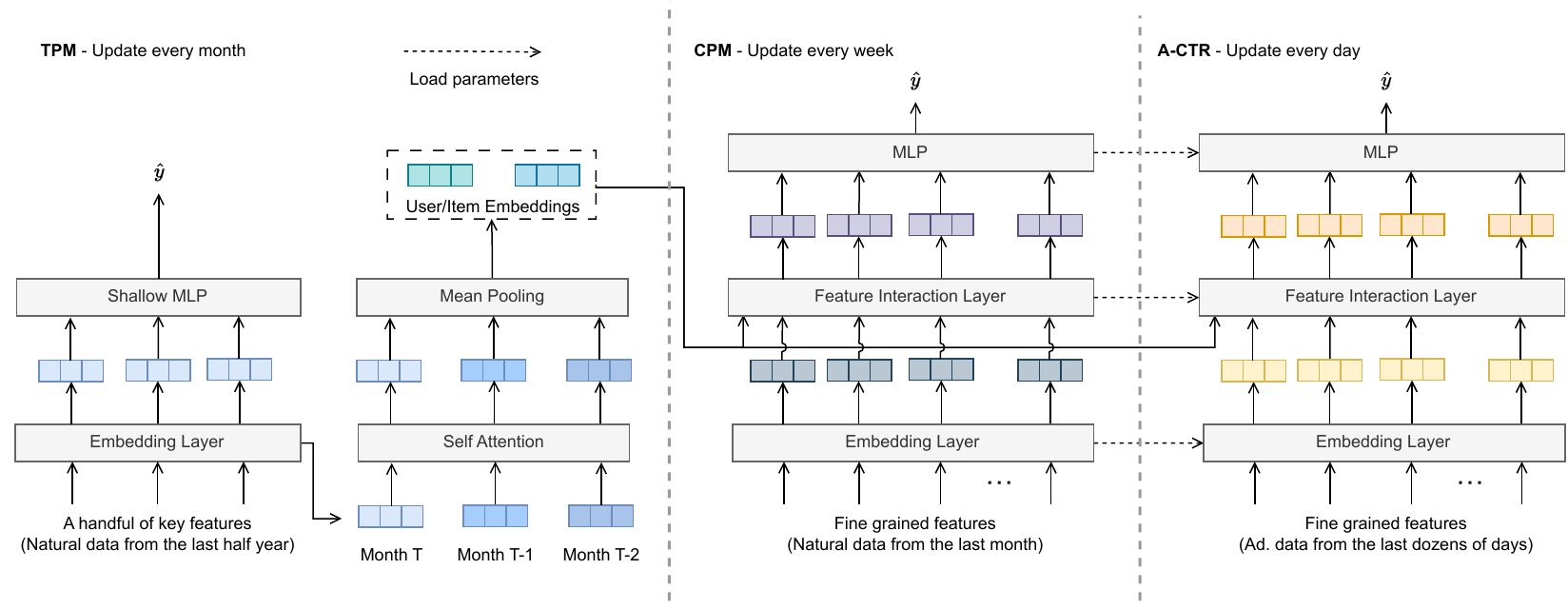}
    \caption{The overall tri-level asynchronous cross-domain framework of E-CDCTR. It consists of the monthly updating Tiny Pre-training Module (TPM), the weekly updating Complete Pre-training Module (CPM), and the daily updating Advertising CTR Model (A-CTR).}
    \label{fig:model}
\end{figure*}


\subsection{Formulation}
The cross-domain CTR prediction is typically understood as the process of knowledge transfer from the source domains to the target domain, leveraging data from all domains to enhance the performance of the CTR prediction model on the target domain. To be more specific, consider a full environment $\mathcal{E}=\{s,t\}$ where $s$ and $t$ stands for source domain and target domain. The input sample pair $(\boldsymbol{X}_i^d, y_i^d)$ are drawn from each domain correspondingly, where domain $d\in\mathcal{E}$ and label $y_i^d\in\{0,1\}$. Notice that $(\boldsymbol{X}_i^s,y_i^s)$ and $(\boldsymbol{X}_i^t,y_i^t)$ are sample pairs from source and target domains respectively. The objective of the CTR prediction task can be summarized as Eq(~\ref{eq:formulation}).
\begin{equation}
    \label{eq:formulation}
    \begin{split}
        &\min_{\theta_{f_t},\theta_g,\theta_{f_s}}\quad\frac{1}{N}\sum_{i=1}^N\mathcal{L}(f_t(\boldsymbol{X}_i^t,y_i^t,g(f_s(\boldsymbol{X}^s,y^s;\theta_{f_s});\theta_g);\theta_{f_t})),\\
        &\mathcal{L}(y,\hat{y})=-ylog(\hat{y})-(1-y)log(1-\hat{y})
    \end{split}
\end{equation}
where $N$ stands for the number of samples in target domain, $\mathcal{L}$ is the cross-entropy loss function. $f_t$ is the target domain CTR model, $g$ is the knowledge transfer model, and $f_s$ is the source domain model. $\theta_{f_t},\theta_g,\theta_{f_s}$ are their parameters respectively.

\subsection{Overall Framework}
The overall framework of E-CDCTR is shown in Figure~\ref{fig:model}, which is a tri-level asynchronous framework consisting of the Tiny Pre-training Model (TPM), the Complete Pre-training Model (CPM), and the Advertising CTR Model (A-CTR). The TPM pre-trains a tiny CTR model which has a shallow Multi-Layer Perceptron (MLP) and takes several basic features as input on the long-term natural data. It aims at providing the historical collaborative filtering signals to the downstream two models. The CPM pre-trains a CTR model sharing the same network structure and input features with the advertising CTR model on short-term natural data. The A-CTR takes the pre-trained model from CPM as initialization parameters and then fine-tunes on the advertising data. They work in concert with each other to make full use of the cross-domain knowledge to optimize the advertising CTR prediction.

\subsection{Tiny Pre-training Model (TPM)}
The online severing CTR model needs to be updated every day in order to fit the latest data distribution. But it can take only the sliding window data from the last dozens of days due to the training cost. However, the user's decision on the platform is not only influenced by instant interest but also affected by long-term interest. The above sliding window-based updating manner will face the challenge of catastrophic forgetting because it can't access the long-term historical data. This problem has a significant adverse impact on the accuracy of CTR estimation, thus we introduce the TPM to mitigate the impact of catastrophic forgetting with high efficiency.

Our main focus is to efficiently preserve collaborative filtering signals from each month so that they can be readily accessed by the downstream model. The most comprehensive collaborative filtering signals stem from the training data itself. Additionally, the model parameters obtained after training also contain a substantial amount of signals. However, accessing either the raw data or the model parameters suffers from the large volume storage cost and training inefficiencies. To address this limitation, we divide the collaborative filtering signals into two parts: user-side and item-side, for separate storage, which greatly enhances the efficiency of signal retrieval. Specifically, we train a tiny CTR model with several basic features on the half-year data in a month-by-month order. During training, user-side and item-side signals are derived from the embeddings corresponding to each user/item at the end of each month. The benefits of splitting and separately storing the signals are evident. Retrieving specific user or item embedding is a mutually independent process, allowing for efficient utilization as feature inputs in the downstream model. Therefore, the TPM module we propose exhibits the following characteristics:
\begin{itemize}
    \item \textbf{Key Features}: TPM input includes only a handful of key features, such as user\_id, item\_id, category\_id, user\_profile, item\_profile, etc., while discarding numerous secondary features.
    \item \textbf{Simple Network Architecture}: TPM has a lightweight architecture that consists of an embedding layer, and a shallow MLP.
    \item \textbf{Monthly Updating}: TPM is updated using data from the past half year on the first day of every month, generating the user/item embeddings for the model training of the current month.
    \item \textbf{Historical Embeddings Storage}: TPM maintains a group of representations comprising user and item embeddings generated over the last three months.
\end{itemize}
When CPM and A-CTR consume features of a certain sample during training, they will retrieve the corresponding three user/item embeddings from the embedding table based on user\_id and item\_id. We concatenate the retrieved user/item embeddings as Eq(~\ref{eq:emb_concat}).
\begin{equation}
    \label{eq:emb_concat}
    e = [emb_1, emb_2, emb_3] \in R^{3*d}
\end{equation}
where $d$ is the embedding dimension and $emb_j$ represents the last $j$-th month's user/item embedding. Before serving as the input of the downstream CTR models, $e$ will undergo self-attention~\cite{vaswani2017attention} to extract contextual information of the CF signals further and be compressed into a single input feature using mean pooling operation. This process can be represented as Eq(~\ref{eq:emb_agg}).
\begin{equation}
    \label{eq:emb_agg}
    \begin{split}
        &e_{SA} = \text{Self\_Attention}(e) \\
        &e_{mean} = \text{Mean\_Pooling}(e_{SA})
    \end{split}
\end{equation}
As a result, the embeddings fed into subsequent models are also divided into user-side $e^u_{mean}$ and item-side $e^i_{mean}$. This division is motivated by the consideration that three embedding tables can significantly increase memory and latency costs during online inference. By employing self-attention to aggregate the three embedding tables, combining them into one embedding table is available after the parameters of the A-CTR are fixed, thus reducing memory and latency costs.
    
\subsection{Complete Pre-training Model (CPM)}
CTR model trained on advertisement data prevalently suffers from data sparsity which means that positive samples take up only a small part of the total samples. But, in industrial recommendation systems, natural data is larger in quantity and denser in CTR. An intuitive idea is to transfer interaction signals from the natural domain to the target advertisement domain through knowledge transfer, aiming to alleviate the training convergence challenge caused by data sparsity. However, the training cost of extensive natural data is commonly unacceptable. To further enhance data utilization, we adopt the pre-train and fine-tune paradigm to better adapt knowledge transfer. 

Utilizing the pre-train and fine-tune paradigm necessitates that the network structure of the pre-trained natural CTR model and fine-tuned advertisement CTR model align with each other. Thus, CPM and A-CTR maintain consistency in network structure. After CPM finishes training on natural data, its parameters are transferred into A-CTR with the aim of offering good initialization for accelerating the convergence of A-CTR. Compared to TPM, CPM is designed with the following characteristics: 
\begin{itemize}
\item \textbf{Complete Features}: CPM's input includes complete features compared to the few selected key ones in TPM, such as user behavior sequences, context information, etc.
\item \textbf{Model Complexity}: CPM comprises behavior sequence models, feature cross models, and a larger MLP layer, thus leading to a more powerful modeling capability.
\item \textbf{Short-Term Training}: CPM utilizes natural traffic data from the last month with a more frequent weekly update.
\item \textbf{Historical Embeddings Consumption}: CPM supplements its input with the user/item embeddings generated by TPM from the last three months. As Figure~\ref{fig:date_arch} illustrates, these embeddings are accessed in a non-overlapping manner between TPM and CPM, meaning that the month of embeddings generated by TPM is the non-overlapping last three months with respect to the month of natural data used for training CPM. This displacement is essential to mitigate the problem of sample re-training, which would degrade performance.
\end{itemize}
Once the finishing of CPM training, its model parameters are used as the initialization for A-CTR. However, it's important to note that the parameters within the batch normalization (BN) are re-initialized. This is due to the differences in data distribution between natural data and advertisement data. The parameters related to data distribution in the deep neural network are primarily encapsulated within BN. Re-initializing these parameters helps refresh A-CTR and further improve the performance. CPM is pre-trained using the binary cross entropy loss.

\subsection{Advertising CTR Model (A-CTR)}
Every day, A-CTR loads the parameters from the latest checkpoint of CPM for initialization and then fine-tunes on the last month's advertisement data consuming the historical embeddings from TPM. The fine-tuning process relies solely on the data from the target advertising domain without the natural source domain. The separation of two phases significantly contributes to an efficient and practical deployment strategy. A-CTR is fine-tuned using the binary cross entropy loss. After the fine-tuning is completed, the A-CTR model is deployed for online serving.


\subsection{Deployment}
\subsubsection{Offline Training}
\begin{figure}[htb]
    \centering
    \includegraphics[width=0.8\linewidth]{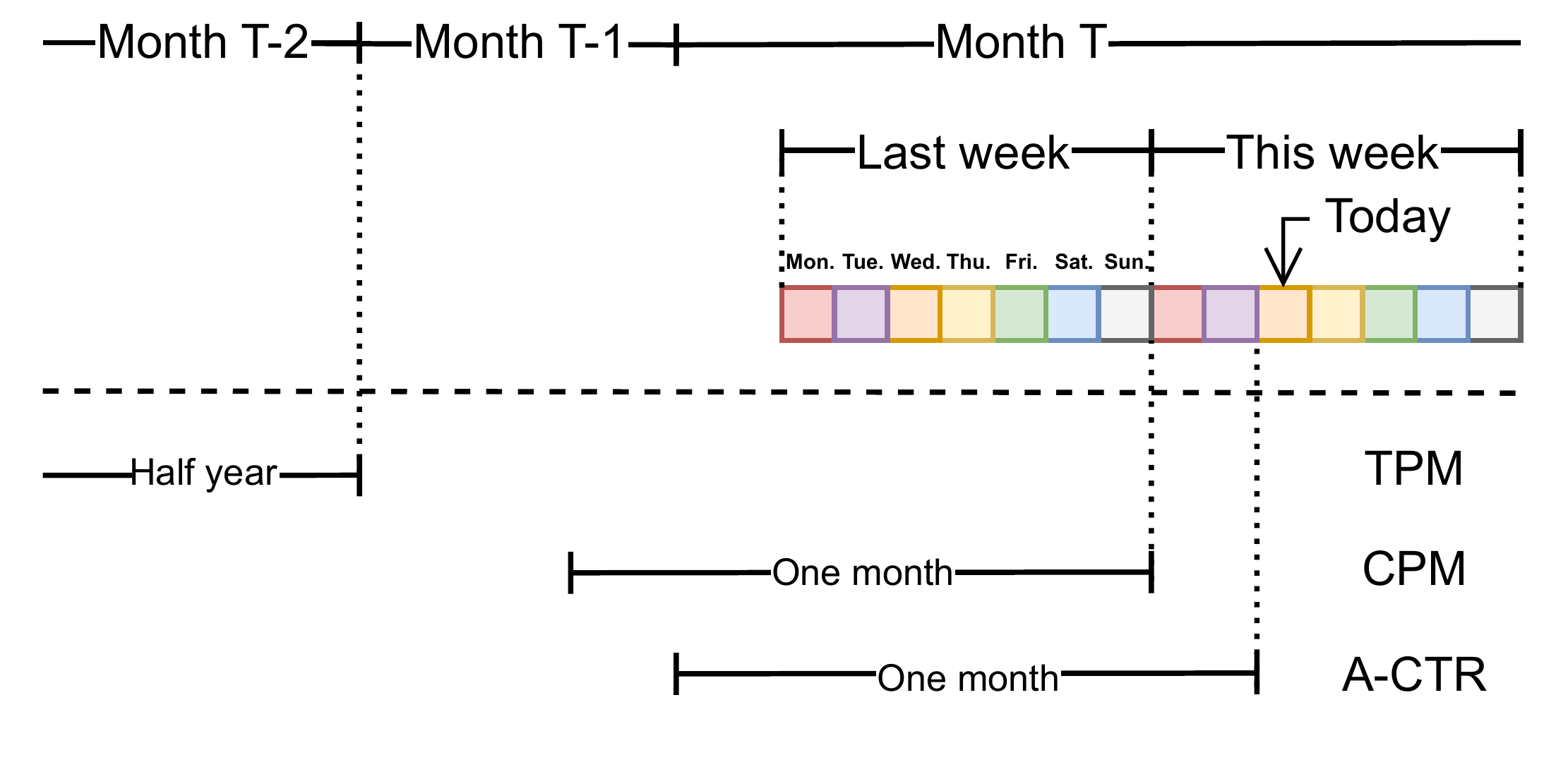}
    \caption{The training data period for three components.}
    \label{fig:date_arch}
\end{figure}
Considering effectiveness and efficiency, we update the three CTR models at different frequencies with different training data periods, which is shown in Figure~\ref{fig:date_arch}. We update the TPM CTR model on the first day of every month as the user/item's long-term characteristics remain stable for one month. For updating, we collect the natural data of the last half year to train the tiny CTR model in a month-by-month order and store user/item embedding tables at the end of each month. We collect data from the last half year instead of the last three months because we aim at utilizing the data of the oldest three months for warming up the model's training, and we only store the user/item embedding tables of the last three months. After obtaining the newly generated embedding tables, we will discard the old ones for saving storage. The CPM updates on Monday of every week using the last month's natural data. The pre-trained parameters will be saved and the checkpoint of last week will be abandoned. Every day, the A-CTR model will load the latest checkpoint from CPM for parameter initialization but abandon the parameters in BN. Then, A-CTR fine-tunes itself using the last month's advertisement data.

\subsubsection{Online Serving}
During the online serving stage, only the A-CTR CTR model together with the user/item embeddings of the last three months will be involved in the inference computation. As we apply the self\_attention to aggregate the user/item embeddings and the parameters of self\_attention are fixed after fine-tuning, we merge the three user/item embeddings into one user/item embedding table using the fixed self\_attention. This merging operation can reduce the storage and latency of the online inference process. 

\section{Experiment Setup}
\subsection{Datasets}
There is no large-scale industrial recommendation system dataset that contains both the natural and advertising impression logs simultaneously, we construct the experimental data with the impression log from the Meituan\footnote{\url{https://www.meituan.com/}} platform for the offline experiment. This dataset consists of two parts: a natural domain dataset and an advertising domain dataset. The natural domain dataset consists of the last seven months' impression logs. The oldest six months of data are used for TPM and the last month's data is applied in the CPM. The collected last month's impression logs of the advertisement domain are used for fine-tuning the A-CTR model. The statistics of all three datasets are shown in Table~\ref{tab:data_stats}. It can be found that the natural domain contains much more impressions than the advertisement domain. This helps alleviate the data sparsity for both the user and the item.

\subsection{Baselines}
To verify the effectiveness, we compare E-CDCTR with the following state-of-the-art cross-domain CTR prediction methods.
\begin{itemize}
    \item \textbf{Target-only} Target-only is the standard paradigm, which is the basic baseline to measure the performance of other methods. It only uses the samples of the advertisement domain to train the target domain's CTR model.
    \item \textbf{CoNet}~\cite{hu2018conet} adds the cross-connections between the source network and the target network through the learnable parameters to achieve dual knowledge transfer between domains. 
    \item \textbf{DARec}~\cite{yuan2019darec} tries to enhance the cross-domain knowledge by learning the shared domain-invariant knowledge with the help of DANN\cite{ganin2015unsupervised}.
    \item \textbf{MiNet}~\cite{ouyang2020minet} extracts the long and short interests from the source domain and takes the interests to assist the user modeling of the target domain.
    \item \textbf{Share-Bottom}~\cite{caruana1997multitask} is a multi-task learning method but is suitable for multi-domain learning. The embedding layer is shared and each domain holds its own MLP network. 
    \item \textbf{Star} introduces a shared MLP network to connect domains and individual batch normalization on the basis of Share-bottom. 
    \item \textbf{MMoE}~\cite{ma2018modeling} utilizes multiple expert networks and gate networks to learn the domain-shared and domain-specific patterns. 
    \item \textbf{KEEP}~\cite{zhang2022keep} stores the user/item embedding at the end of the source domain pre-training. Those embedings will be plugged into the target domain's model serving as extra knowledge. 
\end{itemize}

\begin{table}[tb!]
\renewcommand\arraystretch{0.85}
\caption{Statistics of training datasets for the three components of E-CDCTR, where K means thousand, M represents million, and B donates billion.}
\centering
\label{tab:data_stats}
\resizebox{1.\columnwidth}{!}{
\begin{tabular}{cl|c|c|c|c}
\toprule
\multicolumn{2}{c|}{Datasets}   & \multicolumn{1}{c}{\#Users} & \multicolumn{1}{c}{\#Items} & \multicolumn{1}{c}{\#Features} & \multicolumn{1}{c}{\#Samples} \\
\midrule
\multicolumn{2}{c|}{TPM}           & \multicolumn{1}{c}{192M}             & \multicolumn{1}{c}{16M}            & \multicolumn{1}{c}{19}                & \multicolumn{1}{c}{21B}  \\
\multicolumn{2}{c|}{CPM}           & \multicolumn{1}{c}{192M}             & \multicolumn{1}{c}{16M}            & \multicolumn{1}{c}{292}               & \multicolumn{1}{c}{3.4B}  \\
\multicolumn{2}{c|}{A-CTR}   & \multicolumn{1}{c}{161M}             & \multicolumn{1}{c}{459K}           & \multicolumn{1}{c}{292}               & \multicolumn{1}{c}{2.6B}  \\
\bottomrule 
\end{tabular}
}
\end{table}

\subsection{Evaluation Metric}
The widely used Group AUC (GAUC)~\cite{zhou2018deep} which is consistent with our online performance is for offline assessing. The GAUC measures the weighted average ranking accuracy with respect to users. A higher GAUC indicates better ranking performance for each user. GAUC is calculated as Eq(~\ref{eq:gauc}).
\begin{equation}
    \label{eq:gauc}
    GAUC=\frac{\sum^U_{u=1}\#impr(u)*AUC(u)}{\sum^U_{u=1}\#impr(u)}
\end{equation}
where $U$ represents the number of users, $\#impr(u)$ donates the number of impressions of $u$-th user, and $AUC(u)$ is calculated using the samples of $u$-th user. In an industrial recommendation system, even a slight improvement at \textbf{0.001} level is considered a significant boost~\cite{guo2017deepfm} because it leads to a significant increase in revenue.

\subsection{Implementation Details}
For a fair comparison, the data used for the training model needs to be aligned as far as possible. The last month's natural and advertising data serve for training for the co-training-based methods including CoNet, DARec, MiNet, Share-Bottom, Star, and MMoE. For KEEP, the last seven months' data are for pre-training. We implement E-CDCTR with Tensorflow. The embedding size is $16$ for all fields in all experiments. We train all models using $8$ $80G$ $A100$ GPUs and use Adam~\cite{kingma2014adam} as the optimizer. For TPM, the learning rate is $0.025$ and the batch size is $500,000$. In the CPM and A-CTR model training, the learning rate is $0.0005$ and the batch size is $10,000$. We ran all experiments three times and reported the average result. 

\section{Experiment Results}
\subsection{Overall Performance}
\begin{table}[tb!]
\centering
\caption{Performance of all methods on the industrial datasets. The best result is in boldface and the second best is underlined. * indicates that the difference to the best baseline is statistically significant at 0.01 level~\cite{guo2017deepfm}.}
\label{tab:main_result}
\begin{tabular}{cl|c|c|c|c}
\toprule
& & \multicolumn{1}{c}{GAUC} &\multicolumn{1}{c}{Improv.} \\
\midrule
\multicolumn{2}{c|}{Target-only}        & \multicolumn{1}{c}{0.6564}   & \multicolumn{1}{c}{-} \\
\multicolumn{2}{c|}{Share-Bottom}        & \multicolumn{1}{c}{0.6577}   & \multicolumn{1}{c}{+0.0013} \\
\multicolumn{2}{c|}{CoNet}        & \multicolumn{1}{c}{0.6575}   & \multicolumn{1}{c}{+0.0011} \\
\multicolumn{2}{c|}{DARec}        & \multicolumn{1}{c}{0.6562}   & \multicolumn{1}{c}{-0.0002} \\
\multicolumn{2}{c|}{MiNet}        & \multicolumn{1}{c}{0.6572}   & \multicolumn{1}{c}{+0.0008} \\
\multicolumn{2}{c|}{Star}        & \multicolumn{1}{c}{0.6558}   & \multicolumn{1}{c}{-0.0006} \\
\multicolumn{2}{c|}{MMoE}        & \multicolumn{1}{c}{0.6579}   & \multicolumn{1}{c}{+0.0015} \\ 
\multicolumn{2}{c|}{KEEP}        & \multicolumn{1}{c}{\underline{0.6582}}   & \multicolumn{1}{c}{+0.0018} \\ \midrule
\multicolumn{2}{c|}{E-CDCTR}        & \multicolumn{1}{c}{\textbf{0.6652$^*$}}   & \multicolumn{1}{c}{+0.0088} \\ 
\bottomrule 
\end{tabular}
\end{table}

Table~\ref{tab:main_result} shows the GAUC of all methods and the relative improvement compared to the Target-only. E-CDCTR achieves the best performance among all methods, which indicates its superiority in the cross-domain CTR prediction task. Further, we have some insightful findings from the results.
(1) The proposed E-CDCTR reaches the best performance. Compared with existing cross-domain methods, E-CDCTR can exploit more data from the source domain and achieve the knowledge transfer including the multiple historical user/item embeddings and the pre-trained parameters in an efficient tri-level cross-domain knowledge transfer framework. 
(2) A reasonable knowledge transfer network structure is necessary. The result of Star is worse than Target-only, which demonstrates Star's design will cause a negative transfer between the natural domain and the advertising domain. We think the imbalance ratio of the two datasets and the tightly coupled MLP parameters are the culprits. DARec just holds comparable performance with Target-only. The domain-invariant pattern may be not useful for cross-domain CTR knowledge transfer. 
(3) The other co-training methods outperform Target-only. The result shows that the knowledge of the abundant source domain data can benefit the target domain with proper network structure. On the other side, the in-distinctive performance among them indicates that the training of the natural and advertising CTR models in a unified manner reaches the diminishing marginal returns dilemma.
(4) KEEP outperforms among all baseline methods. The result indicates that user/item embeddings pre-trained on the long-term source data benefit the target domain's learning. We think the improvements also come from alleviating the catastrophic forgetting problem. On the other hand, it also demonstrates that transferring knowledge by pre-trained embeddings is more effective and efficient than co-training methods. 
(5) Our proposed E-CDCTR achieves better performance than KEEP with a slight additional training cost. It is because we make full use of knowledge in the source data through multiple historical user/item embeddings and parameters pre-trained with full features on a complex CTR model, both of which together boost the performance significantly. 

\subsection{Ablation Studies}
\begin{table}[tb!]
\centering
\caption{The result of each component's effect.}
\label{tab:ablation_component}
\begin{tabular}{cl|c|c|c|c}
\toprule
& & \multicolumn{1}{c}{GAUC} &\multicolumn{1}{c}{Improv.} \\
\midrule
\multicolumn{2}{c|}{Target-only}        & \multicolumn{1}{c}{0.6564}   & \multicolumn{1}{c}{-} \\
\multicolumn{2}{c|}{+TPM}        & \multicolumn{1}{c}{0.6597}   & \multicolumn{1}{c}{+0.0033} \\
\multicolumn{2}{c|}{+CPM}        & \multicolumn{1}{c}{0.6620}   & \multicolumn{1}{c}{+0.0056} \\ \midrule
\multicolumn{2}{c|}{E-CDCTR}        & \multicolumn{1}{c}{0.6652}   & \multicolumn{1}{c}{+0.0088} \\
\bottomrule 
\end{tabular}
\end{table}
In this section, we explore the influence of TPM and CPM on the target domain's CTR prediction performance. The ablation studies are derived on the Target-only baseline. We integrate TPM with Target-only by only taking the historical user/item embeddings of the last three months as extra input features. For combining CPM with Target-only, we first pre-train the complex CTR model on the natural data and then fine-tune on the advertisement data, discarding the BN's parameters. 

From Table~\ref{tab:ablation_component}, we can find that both TPM and CPM are beneficial for cross-domain CTR prediction. TPM supplies multiple historical user/item embeddings for A-CTR and achieves promotion by alleviating the catastrophic forgetting problem. CPM boosts the performance by providing a good parameters initialization for A-CTR. CPM is more outstanding than TPM, which indicates that the complete features and complex models are more informative.

\subsection{Analysis of Component Design}
\subsubsection{The Way of Utilizing Samples from Source Domain}
\begin{table}[tb!]
\centering
\caption{GAUC of using source domain data in different ways.}
\label{tab:ablation_sample}
\begin{tabular}{cl|c|c|c|c}
\toprule
& & \multicolumn{1}{c}{GAUC} &\multicolumn{1}{c}{Improv.} \\
\midrule
\multicolumn{2}{c|}{Target-only}        & \multicolumn{1}{c}{0.6564}   & \multicolumn{1}{c}{-} \\
\multicolumn{2}{c|}{Source-only}        & \multicolumn{1}{c}{0.6396}   & \multicolumn{1}{c}{-0.0168} \\
\multicolumn{2}{c|}{Sample Merging}        & \multicolumn{1}{c}{0.6560}   & \multicolumn{1}{c}{-0.0040} \\ \midrule
\multicolumn{2}{c|}{E-CDCTR}        & \multicolumn{1}{c}{0.6652}   & \multicolumn{1}{c}{+0.0088} \\
\bottomrule 
\end{tabular}
\end{table}
In this subsection, we investigate how the method of utilizing samples from the source domain affects the GAUC metric. Herein, we do experiments under two simple and straightforward methods. The Source-only means that we directly train the complex CTR model on last month's natural data. For Sample Merging, we mix the natural and advertisement data of the last month. The mixed data is used for training the complex CTR model. After training, we calculate the GAUC of the trained model on the testing set of the advertisement domain. Note that there is no TPM here. 

The experiment results are shown in Table~\ref{tab:ablation_sample}. We can see that neither of them surpasses the Target-only baseline. Source-only performs terribly because the data distribution between training and testing is inconsistent. Sample Merging is better than Source-only by mixing advertisement data but still worse than Target-only due to the same reason. These results tell the fact it is not trivial to realize positive cross-domain knowledge transfer for the CTR prediction task. 

\subsubsection{Effect of historical embeddings.}
\begin{table}[tb!]
\centering
\caption{Performance of taking different months' historical embeddings.}
\label{tab:ablation_periodicity}
\begin{tabular}{cl|c|c|c|c}
\toprule
& & \multicolumn{1}{c}{GAUC} &\multicolumn{1}{c}{Improv.} \\
\midrule
\multicolumn{2}{c|}{Target-only+CPM}        & \multicolumn{1}{c}{0.6620}   & \multicolumn{1}{c}{-} \\ \midrule
\multicolumn{2}{c|}{One Month}        & \multicolumn{1}{c}{0.6642}   & \multicolumn{1}{c}{+0.0022} \\
\multicolumn{2}{c|}{Two Months}        & \multicolumn{1}{c}{0.6647}   & \multicolumn{1}{c}{+0.0027} \\
\multicolumn{2}{c|}{Three Months (E-CDCTR)}        & \multicolumn{1}{c}{0.6652}   & \multicolumn{1}{c}{+0.0032} \\
\bottomrule 
\end{tabular}
\end{table}
Since we insist on making CPM and A-CTR taking multiple historical user/item embeddings generated by TPM, we conduct experiments to explore its real effect. The experiments are conducted on Target-only with CPM variants. Based on it, we introduce the TPM and increase the historical user/item embedding(s) from the last month to the last three months. Table~\ref{tab:ablation_periodicity} shows that increasing the number of months of historical embeddings gives a better CTR prediction accuracy. However, introducing even more historical embeddings would cause a GPU out-of-memory error in our hardware configuration because the huge number of users/items in Meituan leads to corresponding memory-intensive embedding tables.

\subsubsection{Which part parameters of CPM should be loaded?}
\begin{table}[tb!]
\centering
\caption{Results of loading different part parameters of CPM.}
\label{tab:ablation_parameters}
\begin{tabular}{cl|c|c|c|c}
\toprule
& & \multicolumn{1}{c}{GAUC} &\multicolumn{1}{c}{Improv.} \\
\midrule
\multicolumn{2}{c|}{Target-only+TPM}        & \multicolumn{1}{c}{0.6597}   & \multicolumn{1}{c}{-} \\ \midrule
\multicolumn{2}{c|}{Embeddings}        & \multicolumn{1}{c}{0.6621}   & \multicolumn{1}{c}{+0.0024} \\
\multicolumn{2}{c|}{MLP (w/o BN)}        & \multicolumn{1}{c}{0.6632}   & \multicolumn{1}{c}{+0.0035} \\
\multicolumn{2}{c|}{ALL}        & \multicolumn{1}{c}{0.6645}   & \multicolumn{1}{c}{+0.0048} \\
\multicolumn{2}{c|}{E-CDCTR}        & \multicolumn{1}{c}{0.6652}   & \multicolumn{1}{c}{+0.0055} \\
\bottomrule 
\end{tabular}
\end{table}
In this subsection, we decide to analyze which part parameters should be transferred from the CPM to the A-CTR for initialization. We choose Target-only equipped with the TPM variant as the footstone. According to the network structure of the CTR model, we divide parameters into two parts including embeddings and parameters of MLP. After training the CPM, we take either embeddings or MLP's parameters without BN to initialize the corresponding parameters in A-CTR while the rest parameters of A-CTR are initialized randomly. We also try to load all parameters from CPM including embeddings, parameters of MLP, and BN. The experiment results are shown in Table~\ref{tab:ablation_parameters}. We can see that both the embeddings and parameters of MLP without BN are beneficial for the following A-CTR model. Loading them together into the A-CTR which is the E-CDCTR can boost the A-CTR's performance further. When loading all parameters including the parameters of BN, the result decreases compared to E-CDCTR. The above results demonstrate the embeddings and parameters of MLP without BN containing mostly collaborative filtering knowledge are effective for fine-tuning, while the parameters of BN which memorize the data distribution are harmful to the fine-tuning of A-CTR due to the disagreement of distribution between the natural content and advertising data.

\subsubsection{Hyperparamter: Dimension of Historical Embeddings}
\begin{figure}[htb]
    \centering
    \includegraphics[width=0.8\linewidth]{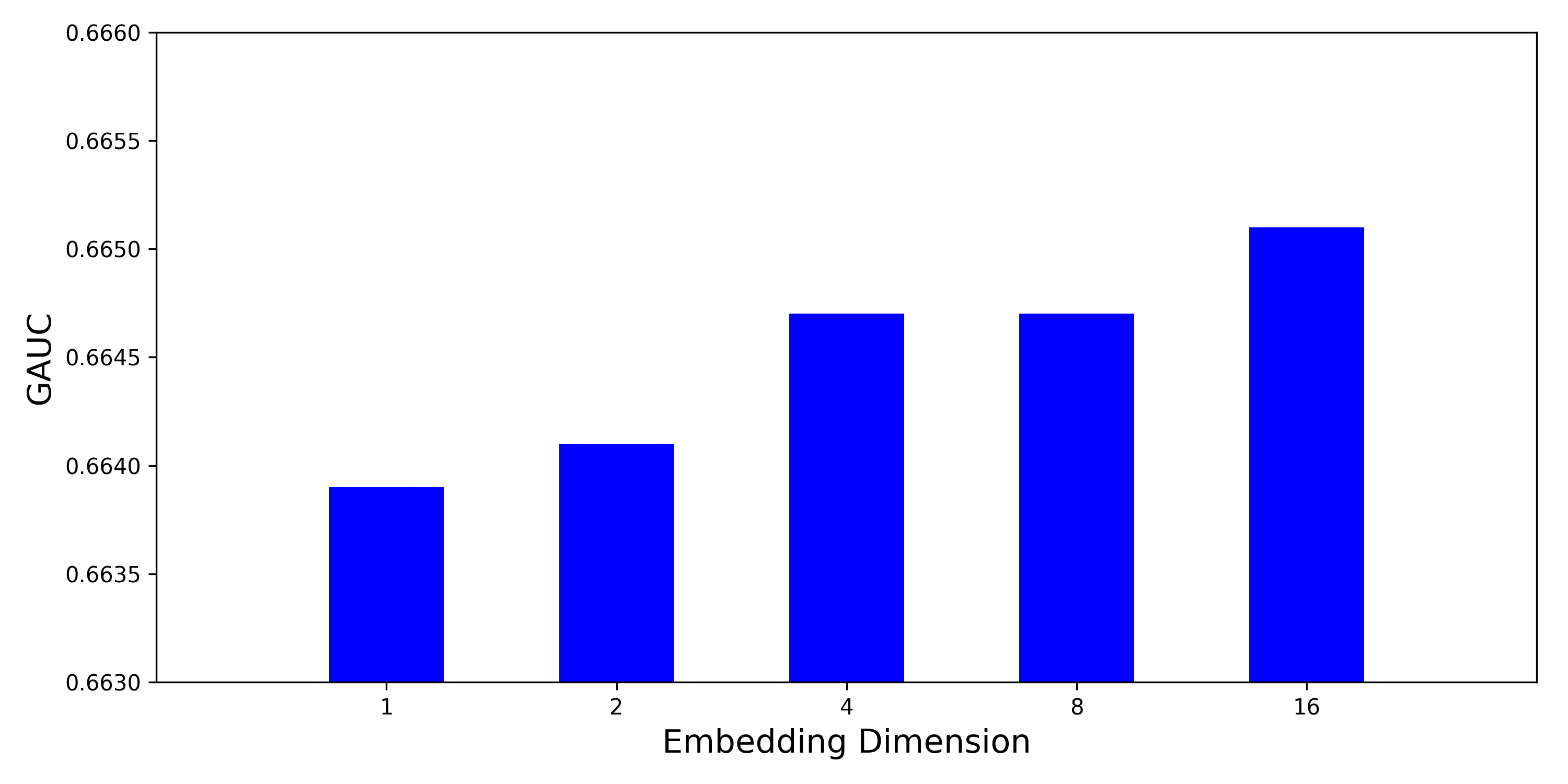}
    \caption{Effect of historical embeddings' dimension.}
    \label{fig:emb_dim}
\end{figure}

Figure~\ref{fig:emb_dim} demonstrates the effect of the historical embedding dimension. In the experiment, we kept all other settings and parameters constant while observing the changes in GAUC by adjusting the embedding dimension. Concluding from Figure~\ref{fig:emb_dim}, it is evident that as the dimension increases, the model shows a significant trend of improved GAUC. This affirms a positive correlation between the model's embedding dimension and its representational capacity. Therefore, given computational resources, opting for larger embedding dimensions contributes to the growth of accuracy metrics for CTR estimations.

\subsection{Online A/B Test}
\begin{table}[h]
    \caption{A/B Test of E-CDCTR compared to Target-only.}\label{tab:online_result}
    \vspace{-0.4cm}
    \setlength{\tabcolsep}{8mm}
    \begin{tabular}{c|c}
    \toprule
    Metrics            & Accumulated Gains   \\ \midrule 
    CTR     & +2.9\% \\ 
    RPM     & +2.1\%  \\ \bottomrule  
    \end{tabular}
\end{table}
We conducted an A/B test in Meituan's online advertising system to measure the benefits of E-CDCTR compared with the online baseline Target-only. The E-CDCTR is allocated with 10\% serving traffic for 30 days from 2023-04 to 2023-05. Table~\ref{tab:online_result} shows the relative promotion of CTR and Revenue Per Mille (RPM). E-CDCTR achieved $2.9\%$ and $2.1\%$ accumulated relative promotion on the CTR and RPM respectively during the A/B test period. This is a significant improvement in the online advertising system and proves the effectiveness of E-CDCTR. 

\section{Conclusion}
We propose the tri-level asynchronous framework E-CDCTR which contains TPM, CPM, and A-CTR to achieve efficient cross-domain knowledge transfer for CTR prediction. TPM focuses on recording long-term collaborative filtering through stored embeddings. CPM aims to provide good initialization points for A-CTR to alleviate the data sparsity problem. A-CTR fine-tunes itself on the advertisement data. We conduct offline/online experiments on the industrial dataset of Meituan to verify the effectiveness of the E-CDCTR.


\bibliographystyle{ACM-Reference-Format}
\bibliography{sample-base}

\appendix

\end{document}